\documentclass{article}
\usepackage{amssymb, latexsym}
\title{Comparing Computational Power}
\author{Udi Boker and Nachum Dershowitz\\School of Computer Science, Tel Aviv University\\Ramat Aviv, Tel Aviv 69978, Israel\\
   E-mail: {\tt\{udiboker,nachumd\}@tau.ac.il}}
\newcommand{\floor}[1]{\left\lfloor#1\right\rfloor}
\newcommand{\rt}[1]{\floor{\sqrt#1}}
\newcommand{\st}{\mathrel{:}}
\newcommand{\Nat}{\ensuremath{\mathbb N}}

\newcommand{\Rec}{{\rm Rec}}
\newcommand{\Prim}{{\rm Prim}}
\newcommand{\RE}{{\rm RE}}
\newcommand{\Part}{{\rm PR}}
\newcommand{\TM}{{\rm TM}}
\newcommand{\ITM}{{\rm ITM}}
\newcommand{\map}[2]{#1\langle#2\rangle}
\newcommand{\dom}{{\rm dom}\;}
\newcommand{\rng}{{\rm rng}\;}
\newcommand{\subject}[1]{\vspace*{3 mm} \noindent \emph{#1}~}

\newtheorem{theorem}{Theorem}
\newtheorem{corollary}{Corollary}
\newtheorem{definition}{Definition}
\newtheorem{proposition}{Proposition}
\newtheorem{lemma}{Lemma}
\newtheorem{example}{Example}
\newenvironment{proof}{\paragraph{Proof.}}{$\Box$}
\begin{document}
\sloppy
\date{April 2005}
\maketitle

\begin{quote}\raggedleft
\emph{All models are wrong but some are useful. } \\[2ex]
---George E.\ P.\ Box,\\ ``Robustness in the strategy of\\scientific model
building'' (1979)
\end{quote}

\begin{abstract}
It is common practice to compare the computational power of
different models of computation.
For example, the recursive functions are strictly more
powerful than the primitive recursive functions,
because the latter are a proper subset of the former
(which includes Ackermann's function).
Side-by-side with this ``containment'' method of measuring power,
it is standard to use an approach based on ``simulation''.
For example, one says that the (untyped) lambda calculus is as powerful---computationally
speaking---as the partial recursive functions, because
the lambda calculus can simulate all partial recursive functions by encoding
the natural numbers as Church numerals.

The problem is that unbridled use of these two ways of comparing
power allows one to show that some computational models are
\textit{strictly} stronger than themselves!
We argue that a better definition is that model $A$ is strictly stronger
than $B$ if $A$ can simulate $B$ via some encoding,
whereas $B$ cannot simulate $A$ under \textit{any} encoding.
We then show that the recursive functions are strictly stronger in this sense
than the primitive recursive.
We also prove that the recursive functions, partial recursive
functions, and Turing machines are ``complete'', in the sense that
no injective encoding can make them equivalent to any ``hypercomputational'' model.
\end{abstract}

\section{Introduction}

Our overall goal is to formalize the comparison of computational
models. We seek a robust definition of relative power that
does not itself depend on the notion of computability. It should allow
one to compare arbitrary models over arbitrary domains via a
quasi-ordering that successfully captures the intuitive concept of
computational strength. We want to be able
to prove statements like ``analogue machines are strictly more
powerful than digital devices'', even though the two models
operate over domains of different cardinalities.

Since we are only
interested here in the extensional quality of a computational model (the set of
functions that it computes), not complexity-based comparison or
step-by-step simulation,
we use the term ``model''
for any set of partial functions,
and ignore all the ``mechanistic'' aspects.

\subsection{The Standard Comparison Method}\label{Sec:StandardMethod}
There are basically two standard methods, Approaches C and S below, by which models have
been compared over the years. These two approaches have been
used in the literature in conjunction with each other; thus,
they need to work in harmony. That is, if models $A$ and $A'$
are deemed equivalent according to approach C, while $A'$ is shown to be
stronger than $B$ by approach S, we expect that it is legitimate
to infer that $A$ is also stronger than $B$.

\subject{Approach C (Containment).} Normally, one would say that a
model $A$ is at least as powerful as $B$ if all (partial)
functions computed by $B$ are also computed by $A$. If $A$ allows
\emph{more} functions than $B$, then it is standard to claim that
$A$ is \emph{strictly} stronger. For example, general recursion
(\Rec) is more powerful than primitive recursion (\Prim) (e.g.\
\cite[p.\ 92]{Tourlakis}), and inductive Turing machines are more
powerful than Turing machines \cite[p.\ 86]{Burgin}.

\subject{Approach S (Simulation).} The above definition does not
work, however, when models use different data structures
(representations). Instead, $A$ is deemed at least as powerful as
$B$ if $A$ can {\em simulate} every function computable by $B$.
Specifically, the simulation is obtained by requiring an injective
encoding $\rho$ from the domain of $B$ to that of $A$, such that
for every function $g$ computed by $B$ we have $g=\rho^{-1}\circ
f\circ\rho$ for some function $f$ computed by $A$, in which case
$A$ is said to be at least as powerful as $B$. See, for example,
\cite[p.\ 27]{Rogers}, \cite[p.\ 24]{Cutland}, or \cite[p.\
30]{Sommerhalder}:
\begin{quote}
Computability relative to a coding is the basic concept in
comparing the power of computation models.\ldots\ The
computational power of the model is represented by the extension
of the set of all functions computable according to the model.
Thus, we can compare the power of computation models using the
concept `incorporation relative to some suitable coding'.
\end{quote}

\subject{Equivalence.} To show that two models are of equivalent
power by the simulation method, one needs to find two injections, each showing that every
function computed by one can be simulated by the other. For example,
Turing machines ($\TM$), the untyped lambda calculus ($\rm\Lambda$),
and the partial recursive functions (\Part) were all shown to be of
equal computational power, in the seminal work of Church \cite{Church}, Kleene
\cite{Kleene} and Turing \cite{Turing}.

\begin{figure}
\hspace*{-3cm}
\begin{picture}(160,170)(-170,-70)
\put(0,0){\oval(40,20)} \put(-10,-3){$\Prim$}
\put(0,0){\oval(80,60)} \put(-20,17){$\Rec \sim \TM$}
\put(0,0){\oval(120,100)} \put(-20,37){$\Part \sim \RE$}
\put(-8,57){$\ITM$} \small \put(-20,-23){Recursive}
\put(-33,-43){Partial Recursive} \put(0,0){\oval(160,140)}
\put(-36,-63){Hypercomputation} \normalsize \put(120,30){$\ITM =$
Inductive Turing Machine} \put(120,15){$\Part$ = Partial
Recursion}\put(120,0){$\RE$ = Recursive enumerable}
\put(120,-15){$\Rec$ = General Recursion} \put(120,-30){$\TM$ =
Turing Machine}\put(120,-45){$\Prim$ = Primitive Recursion}
\end{picture}
\caption{Computational Power Hierarchy}\label{fig:PowerHierarchy}
\end{figure}
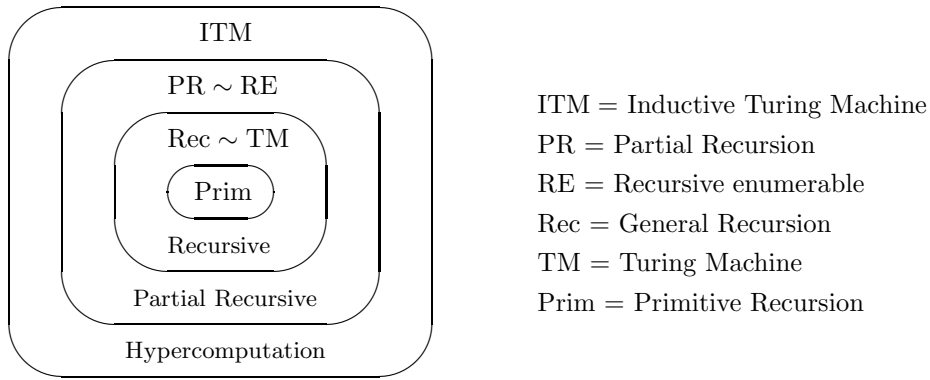

\subject{More Powerful.} To show that model $A$ is strictly more
powerful than model $B$, one normally shows that  $A$ is at least as powerful as
some model $A'$
that comprises more functions than $B$ ($A' \supsetneq B$).  (See, for example,
\cite{Hava}.)
Figure~\ref{fig:PowerHierarchy}
illustrates this standard conception, according to which Turing
machines are considered strictly more powerful than primitive
recursion, since $\TM$ is equivalent to $\Rec$---by simulation, and $\Rec$ is
strictly more powerful than $\Prim$---by containment.

\subsection{The Problem}
Unfortunately,
it turns out that these two approaches, which form the
standard method of comparing computational power, are actually
incompatible.  We provide examples in Section~\ref{sec:submodels}
of cases in which a model $A$ is strictly more powerful than $B$
by the first approach, whereas $B$ is at least as powerful as $A$
by the second.  It follows that the combination of these two
standard approaches allows for models to be strictly stronger than
themselves!

Specifically, in Example~\ref{exm:R2} below, we describe a model
that is a proper subset of the recursive functions, but can, nevertheless, simulate all
of them. This raises the
question whether
it could possibly also be the case that the primitive recursive functions are of
equivalent power to Turing machines, via some ``wild'' simulation.
Could it be
that the recursive functions are of equivalent computational power
to some proper superset, containing non-recursive functions?

\subsection{The Suggested Solution}
We begin (in Definition~\ref{def:ComputationalPower} below) with the
basic comparison notion ``as powerful as'' ($\succsim$), using the
simulation approach (Approach $S$), which naturally extends
containment (Approach C) to models operating over different domains.
(If a model $A$ is as powerful as a $B$ by Approach C, then it is
also as powerful as $B$ by Approach S.) Then the ``strictly more
powerful'' partial ordering ($\succ$) is derived from the
quasi-ordering $\succsim$ by saying that $A \succ B$ if $A \succsim
B$ but not $B \succsim A$, in other words, only when there is no
injection via which $B$ can simulate $A$.




To
compare models operating over different domains requires some
sort of mapping between the domains.  A possible alternative might be
to require a domain mapping that is not only injective but that also possesses additional
properties, like being surjective. It turns out
that bijective mappings not only cannot provide a sufficiently general comparison notion,
but would be limited to permutations with bounded cycles (Theorem
~\ref{thm:BoundedCycles}).

One is tempted
to consider ``well-defined'' only those computational models
that cannot be shown by simulation to be of equivalent power to any proper superset of functions.
We call such a model ``complete''
(Definition~\ref{def:complete}). The question then is:
Are the classic models, such as Turing machines,
well-defined? In Section~\ref{sec:Completeness}, we show that
general recursive functions, partial recursive functions, and
Turing machines are indeed all complete models in our sense
(Theorems~\ref{thm:RecIsComplete}, \ref{thm:PartIsComplete}, and
\ref{thm:TMComplete}).  Accordingly, we obtain a
criterion by which to verify that a model operating over a
denumerable domain is hypercomputational
(Corollary~\ref{cor:hyper}).

\section{Definitions}

We consider only deterministic computational models; hence, we
deal with partial functions.  To simplify the development, we will
assume for now that the domain and range of functions are
identical, except that the range is extended with $\bot$,
representing ``undefined''.

Two partial functions ($f$ and $g$) over the same domain ($D$) are
deemed \emph{(semantically} or \emph{extensionally}) \emph{equal}
(and denoted simply $f=g$) if they are defined for exactly the
same elements of the domain ($f(x)=\bot$ iff $g(x)=\bot$ for all
$x \in D$) and have the same value whenever they are both defined
($f(x)=g(x)$ if $f(x)\neq\bot$, for all $x \in D$).


\begin{definition}[Model of Computation]\label{def:ComputationalModel}
Let $D$ be an arbitrary domain (any set of elements). A \emph{model of computation over}
$D$ is any set of functions $f:D\to D\cup\{\bot\}$. We write
\emph{$\dom A$} for the domain over which model $A$ operates.
\end{definition}

Since models are sets: When $A\subsetneq B$, for models $A$ and
$B$ over the same domain, we say that $A$ is a \emph{submodel} of
$B$ and, likewise, that $B$ is a \emph{supermodel} of $A$.
Moreover, whenever we speak of $A\subseteq B$, we mean to also
imply that the two models operate over the same domain.

To deal with models operating over different domains it is
incumbent to map the domain of one model to that of the other. Let
$\rho:\dom B\cup\{\bot\} \to \dom A\cup\{\bot\}$ be an injective
encoding. Then $\rho\circ M=\{\rho\circ f\st f\in M\}$ and
$M\circ\rho=\{f\circ \rho\st f\in M\}$, for any relation $\rho$
and set of functions $M$. Additionally, we insist that
$\rho(y)=\bot$ iff $y=\bot$.

\begin{definition}[Simulation]\label{def:simulation}
Model $A$ \emph{simulates} model $B$ \emph{via} injection
$\rho:\dom{B}\to\dom{A}$, denoted $A\succsim_\rho B$, if
$\rho\circ B \subseteq A\circ\rho$.
\end{definition}

\noindent This is the notion of ``incorporated'' used in \cite[p.\
29]{Sommerhalder}.

As a degenerate case, with the identity encoding $\iota$ ($\lambda
n.n$), we have $A\succsim_\iota B$ iff $A\supseteq B$. Approach C
of comparison (see the introduction) uses this simple relation.

Approach S is embodied in the following:

\begin{definition}[Computational
Power]\label{def:ComputationalPower}\
\begin{enumerate}
\item
Model $A$ is \emph{(computationally) at least as powerful} as
model $B$, denoted $A\succsim B$, if there is an injection $\rho$
such that $A\succsim_\rho B$.
\item
Model $A$ is \emph{(computationally) more powerful} than $B$,
denoted $A\succ B$, if $A\succsim B$ but $B\not\succsim A$.
\item
Models $A$ and $B$ are \emph{computationally equivalent} if
$A\succsim B\succsim A$, in which case we write $A\sim B$.
\end{enumerate}
\end{definition}

\begin{proposition}
The computational power relation $\succsim$ between models is a
quasi-order. Computational equivalence $\sim$ is an equivalence
relation.
\end{proposition}

Transitivity of $\succsim$ is because the composition of
injections is an injection.

\begin{example}
Turing machines $(\TM)$ simulate the recursive functions $(\Rec)$
via a unary representation of the natural numbers.
\end{example}

\begin{example}
The (untyped) $\lambda$-calculus $(\rm\Lambda)$ is equivalent to the
partial recursive functions $(\Part)$ via Church numerals, on the
one hand, and via G\"{o}delization, on the other.
\end{example}

Since the domain encoding $\rho$ implies, by the simulation
definition, a function mapping, we can extend $\rho$ to functions
and models, as follows:

\begin{definition}[Function Mappings]
An encoding $\rho:\dom B\to\dom A$ induces the following mappings:
\begin{eqnarray*}
\rho(g) &=& \rho\circ g\circ\rho^{-1}\\
\map\rho{f} &=& \rho^{-1}\circ f\circ\rho
\end{eqnarray*}
from $B$ to $A$ and from $A$ to $B$, respectively. These extend to
sets of functions in the usual manner:
\begin{eqnarray*}
\rho(M) &=& \{\rho(g)\st g\in M\}\\
\map\rho{M} &=& \{\map\rho{f}\st f\in M\}\;.
\end{eqnarray*}
\end{definition}

Note that any function $f$, such that
$f\upharpoonright_{\rng\rho}=\rho(g)\upharpoonright_{\rng\rho}$,
simulates $g$ via $\rho$, while $\map\rho f$ is the only function
simulated by $f$. The model $\rho(M)$ is minimal (with respect to
the restriction of the domain to $\rng\rho$) among those that
simulate $M$ via $\rho$, and $\map\rho M$ is the maximal model
simulated by $M$ (see Lemma~\ref{lem:max} below).

\begin{definition}[Strong Equivalence]\label{def:StrongEquivalence}
Model $A$ is \emph{strongly equivalent} to model $B$, denoted $A
\simeq B$, if there are bijections $\pi$ and $\tau$ such that
$A\succsim_\pi B \succsim_\tau A$.
\end{definition}

\begin{definition}[Isomorphism]\label{def:Isomorphism}
Model $A$ is \emph{isomorphic} to model $B$, denoted $A \equiv B$,
if there is a bijection $\pi$ such that $A\succsim_\pi B
\succsim_{\pi^{-1}} A$.
\end{definition}

\begin{example}
Lisp with only pure lists as data is isomorphic to the partial
recursive functions via the G\"{o}del pairing function:
$\pi(\mbox{\bf nil})=0$; $\pi(\mbox{\bf cons}(x,y))=
2^{\pi(x)}(2\pi(y)+1)$.
\end{example}

When $\pi$ is recursive, one may speak of \textit{recursively
isomorphism}: function $f$ is recursively isomorphic to $g$ if
there is a recursive permutation $\pi$, such that $f = \pi^{-1}
\circ g \circ \pi$ \cite[pp.\ 52--53]{Rogers}. Moreover: ``A
property of a $k$-ary relations on $\Nat$ is \textit{recursively
invariant} if, whenever a relation $R$ possesses the property, so
does $g(R)$ for all $g\in{\cal G}^*$'' \cite[p.\ 52]{Rogers},
where ${\cal G}^*$ are the recursive permutations of $\Nat$. Thus,
one may claim: ``[Recursion] theory essentially studies \ldots\
those properties of sets and functions which remain invariant
under recursive permutations. For example, recursiveness, r.e.-ness,
$m$-completeness are such invariants'' \cite[p.\ 333]{Tourlakis}.

\section{Equivalent Submodels}\label{sec:submodels}

Unfortunately, the above standard definition of ``simulates''
(Approach S, Definition~\ref{def:simulation}) allows for the
possibility that a model is equivalent to its supermodel.

\begin{example}\label{exm:R2}
The set of ``even'' recursive functions ($R_2$) is of equivalent
power to the set of all recursive functions. Define:
\begin{eqnarray*}
R_2 &=& \left\{\lambda n. \left\{\begin{array}{ll}
2f(n/2)&\mbox{~$n$ is even}\\
n&\mbox{~otherwise}\end{array}\right\}\st f \in \Rec \right\}
\end{eqnarray*}
We have that $R_2 \succsim_{\lambda n.2n} \Rec$.
\end{example}

Furthermore, it leads to situations where $A \succ B \succ A$ for
models $A,B$. For example, the set of ``odd'' recursive functions
($R_1$, defined analogously) is of equivalent power to the set of
all recursive functions, by the same argument as above. We have
that, $R_1 \succsim \Rec \supsetneq R_2 \succsim \Rec \supsetneq
R_1$, thus $R_1 \succ R_2 \succ R_1$. Thus, the standard
comparison method (Section~\ref{Sec:StandardMethod}) is
ill-defined.

It turns out that the equivalence of a model and its supermodel is
possible even when the encoding $\rho$ is a bijection and the model
is closed under functional composition. Hence, a model might be
isomorphic to a supermodel of itself.

\begin{definition}[Narrow Permutations]
A permutation $\pi:D\to D$ is \emph{narrow} if there is a constant
$k \in \Nat$, such that $\pi^k(x) = x$, for every $x\in D$.
\end{definition}

\begin{theorem}\label{thm:BoundedCycles}
For every encoding $\rho:D\to D$, there are models $A$ and $B$,
such that $A\subsetneq B \precsim_\rho A$, iff $\rho$ is a
non-narrow permutation.
\end{theorem}

\begin{proof}
Suppose that $\pi$ is a permutation with narrow cycles bounded by $k$.
Assume $A \succsim_\pi B \supseteq A$. There is, by assumption, a
function $f\in A$, for every function $g\in B$, such that $g =
\pi^{-1} \circ f \circ \pi$. Since $f\in B$, there is, by
induction, a function $f_k \in A$, such that $g = \pi^{-k} \circ
f_k \circ \pi^k = f_k$. Therefore, $B = A$.

For the other direction, we must consider three cases: (1)
non-surjective encodings; (2) surjective encodings that are not
injective; (3) bijections with no bound on the length of their
cycles. We can prove each case by constructing a computational
model that is strongly equivalent to a supermodel of itself via
the given encoding.

We provide here only a specific instance of case (3); the full
proof is a generalization of the argument.

Let $K$ be a set of ``basic functions'' over \Nat, containing all
the constant functions $\kappa_k$ ($\lambda n.k$), plus the
identity, $\iota$. We present two models, $A$ and $B$, that both
contain the basic functions and are closed under function
composition, such that the smaller one ($B$) simulates every
function of the infinitely larger one ($A$).

Imagine the natural numbers arranged in a triangular array:

\begin{center}
\begin{tabular}{r|rrrrrrrrr}
\bf 0&0\\
\bf 1&1&2&3\\
\bf 2&4&5&6&7&8\\
\bf 3&9&10&11&12&13&14&15\\
\bf 4&16&\multicolumn{2}{c}{\ldots}\\
\vdots&\multicolumn{3}{r}{$\ddots$}\\\hline &\bf 0&\bf 1&\bf 2&\bf
3&\bf 4&\bf 5&\bf 6&\ldots
\end{tabular}
\end{center}

\noindent Now, define the following computable functions:
\begin{eqnarray*}
f_{i,j}(n) &=& \left(\rt{n}+i\right)^2+j\bmod\left(2\rt{n}+2i+1\right)\\
g_i(n) &=& f_{i,0}(n) ~=~ \left(\rt{n}+i\right)^2\!.
\end{eqnarray*}
If $n$ is located on row $m$, then $f_{i,j}(n)$ is the number in
row $n+i$ and column $j$, while $g_i(n)$ is the first number in
row $n+i$.

Consider the following sets of functions:
\begin{eqnarray*}
F &=& \left\{f_{i,j}\st i,j>0\right\}\\
G &=& \left\{g_i\st i>0\right\}.
\end{eqnarray*}
Note that $F$ and $G$ are disjoint, since for every $i, j>0$ and
$n>j^2$, $f_{i-1,j}(n) < g_i(n) < f_{i,j}(n)$.

Define:
\begin{eqnarray*}
B &=& K\cup F\\
A &=& K\cup F\cup G\,.
\end{eqnarray*}
Thus, $A$ has functions to jump anywhere in subsequent rows, while
$B \subsetneq A$ is missing infinitely many functions $g_i$ for
getting to the first position of subsequent rows. Since, for
$i+k>0$,
\begin{eqnarray*}
f_{i,j}\circ f_{k,\ell} &=& f_{i+k,j}\;,
\end{eqnarray*}
it follows that both $F$ and $G$ are closed under composition, as
is their union $F\cup G$, from which it follows that $A$ and $B$
are also closed.

There exists a (computable) permutation $\pi$ of the naturals
\Nat, such that $B \succsim_\pi A$:
\begin{eqnarray*}
\pi(n) &=& f_{0,n-\rt{n}^2+1}\\\nonumber &=&
\rt{n}^2+\left(n-\rt{n}^2+1\right)\bmod\left(2\rt{n}+1\right),
\end{eqnarray*}
mapping numbers to their successor $n+1$, but wrapping around
before each square. That is, $\pi$ has the following unbounded cycles:
\begin{eqnarray*}
\pi &=& \{(0),\;(1\,2\,3),\;(4\,5\,\ldots\,8),\;\ldots\}.
\end{eqnarray*}

It remains to show that for all $f\in A= K\cup F\cup G$, we have
$\pi(f)\in B= K\cup F$.  The following can all be verified:
\begin{eqnarray*}
\pi(\iota) &=& \iota\in K\subseteq B\\
\pi(\kappa_k) &=& \kappa_{\pi(k)}\in K \subseteq B\\
\pi(f_{i,j}) &=& f_{i,j+1}\in B, \mbox{ for $i>0$, $j\geq 0$}\,.
\end{eqnarray*}
\end{proof}

\begin{corollary}
There are models isomorphic to supermodels of themselves.
\end{corollary}

\section{Comparisons}\label{sec:positive}

One can categorize the maximal model that can be simulated, as
follows:

\begin{lemma}\label{lem:max}
For all models $A$ and $B$, $A\succsim_\rho B$ iff
$B\subseteq\map\rho A$.
\end{lemma}

\begin{proof}
We have $B\subseteq\map\rho A$ iff for every $g\in B$ there is
$f\in A$, such that $g = \rho^{-1} \circ f \circ \rho$. This is
the same as requiring that for every $g\in B$ there is an $f\in
A$, such that $\rho \circ g = \rho \circ \rho^{-1} \circ f \circ
\rho = f \circ \rho$, that is, $A\succsim_\rho B$.
\end{proof}

By the same argument:

\begin{lemma}
For all models $A$ and $B$ and bijections $\pi$, $A\succsim_\pi B$
iff $A\supseteq\pi(B)$.
\end{lemma}

\begin{corollary}
For all models $A$ and injections $\rho$, $A\succsim\map\rho A$.
\end{corollary}

\begin{corollary}\label{cor:equiv}
For all models $A$ and bijections $\pi$, $A\simeq \pi(A)$.
\end{corollary}

Clearly, $\map\pi A=\pi^{-1}(A)$.

\begin{lemma}\label{lem:ProperSubsetBijection}
For all models $A$ and $B$ and bijections $\pi$, $A \subsetneq B$
implies that $\pi(A) \subsetneq \pi(B)$ and $\map\pi{A} \subsetneq
\map\pi{B}$.
\end{lemma}

\begin{proof}
Since $\pi$ is a bijection, it follows that $\pi(M)$ is an
injection (i.e. every function of M is simulated by exactly one
function via $\pi$). Therefore, $\pi(B\setminus A) \cap \pi(A)=
\map\pi{B\setminus A} \cap \map\pi{A}=\emptyset$.
\end{proof}

\begin{lemma}\label{lem:StrongEquiv}
If $A \simeq B \subsetneq C$, for models $A,B,C$, then there is a
model $D \supsetneq A$, such that $C \simeq D$.
\end{lemma}

\begin{proof}
Suppose $B\succsim_\pi A$ for bijection $\pi$. Thus,
$A\subseteq\map\pi{B}$. Let $D=\map\pi{C}$, for which we have $C
\simeq D$. Since $B \subsetneq C$, it follows that $A\subseteq
\map\pi{B} \subsetneq \map\pi{C}=D$.
\end{proof}

\begin{theorem}\label{thm:PrimIsWeak}
The primitive recursive functions, \Prim, are strictly weaker than
the recursive functions.
\end{theorem}

\begin{proof}
Clearly, $\Rec\succsim_\iota\Prim$. So, assume, on the contrary,
that $\Prim\succsim_\rho\Rec$. Let $S\in\Rec$ be the successor
function. There is, by assumption, a function $S'\in\Prim$ such
that $S'\circ\rho=\rho\circ S$. Since $\rho(0)$ is some constant
and $\rho(S(n)) = S'(\rho(n))$, we have that $\rho\in\Prim$. Since
$\rho$ is a recursive injection, it follows that $\rho^{-1}$ is
partial recursive. Define the recursive function $h(n)=
\rho(\min_i\{ \rho(i) > ack(n,n)\})$, where $ack$ is Ackermann's
function. Since $\lambda n.ack(n,n)$ grows faster than any
primitive recursive function and $h(n) > ack(n,n)$, it follows
that $h \notin \Prim$. Since $\rng h\subseteq \rng\rho$, it
follows that $t = \rho^{-1} \circ h \in\Rec$. Thus, there is a
function $t' \in \Prim$, such that $t' \circ \rho = \rho \circ t =
\rho \circ \rho^{-1} \circ h = h$. We have arrived at a
contradiction: on the one hand, $t' \circ \rho \in \Prim$, while,
on the other hand, $ h \notin \Prim$.
\end{proof}

\section{Completeness}\label{sec:Completeness}

As shown in Section~\ref{sec:submodels}, a model can be of
equivalent power to its supermodel. There are, however, models
that are not susceptible to such an anomaly.

\begin{definition}[Complete]\label{def:complete}
A model is \emph{complete} if it is not of equivalent power to any
of its supermodels.  That is, $A$ is complete if $A\succsim
B\supseteq A$ implies $A=B$ for all $B$.
\end{definition}

\begin{theorem}\
\begin{enumerate}
\item
Isomorphism of models implies their strong equivalence.
\item\label{2}
Strong equivalence of complete models implies their isomorphism.
\end{enumerate}
\end{theorem}

\begin{proof}
The first statement is trivial. For the second, assume $A
\succsim_\pi B \succsim_\tau A$ for bijections $\pi,\tau$. If
$\map\pi{A} \subsetneq B$, then, by
Lemma~\ref{lem:ProperSubsetBijection}, $\map\tau{\map\pi{A}}
\subsetneq A$, which contradicts the completeness of $A$. Thus $B =
\map\pi{A}$, and therefore, $A = \map{\pi^{-1}}{B}$.
\end{proof}

\begin{lemma}\label{lem:CompleteStrongEquiv}
If model $A$ is complete and $A \succsim_{\rho} B \succsim_{\pi} A$,
for model $B$, injection $\rho$ and bijection $\pi$, then $A$ and
$B$ are strongly equivalent models.
\end{lemma}

\begin{proof}
Suppose $A$ is complete, and $A \succsim_{\rho} B\succsim_{\pi} A$
for injection $\rho$ and bijection $\pi$. It follows that
$\map\pi{B} = A' \supseteq A$. Thus, $A \succsim_{\rho}
B\succsim_{\pi} A' \supseteq A$. Therefore, from the completeness of
$A$, it follows that $A' = A$. Hence, $A' \succsim_{-\pi} B$, and
$A$ and $B$ are strongly equivalent models.
\end{proof}

\begin{theorem}\label{thm:StrongEquivComplete}
If $A$ and $B$ are strongly equivalent models, then $A$ is complete
iff $B$ is.
\end{theorem}

\begin{proof}
Suppose that $A$ is complete and $A \simeq B\subsetneq C$. By
Lemma~\ref{lem:StrongEquiv}, $C \simeq D\supsetneq A$ for some $D$.
Were $B\succsim C$, then $A\simeq B \succsim C\simeq D$,
contradicting the completeness of $A$. Hence, $B$ is also complete.
\end{proof}

\begin{theorem}\label{thm:ABC}
If model $A$ is complete and $A \simeq B \subsetneq C$, for models
$B,C$, then $C \succ A$.
\end{theorem}

\begin{proof}
If $A \simeq B \subsetneq C$, then $C\succsim B \succsim A$. And, by
the previous theorem, if $A$ is complete, then so is $B$; hence
$B\not\succsim C$ and also $A\not\succsim C$. Hence, $C \succ A$.
\end{proof}

We turn now to specific computational models.

\begin{definition}[Hypercomputational Model]\label{def:Hypercomputational}
Model $M$ is \emph{hypercomputational} if there is an injection
$\rho$, such that $\map\rho{M} \supsetneq \Rec$.
\end{definition}

\begin{theorem}\label{thm:RecIsComplete}
The recursive functions \Rec\ are complete. That is, they cannot
simulate any hypercomputational model.
\end{theorem}

\begin{proof}
Assume $\Rec\succsim_\rho M\supseteq\Rec$ and let $S\in M$ be the
successor function. Analogously to the proof of
Theorem~\ref{thm:PrimIsWeak}, $\rho\in\Rec$ and
$\rho^{-1}\in\Part$. For every $f\in M$, there is an $f'\in\Rec$,
such that $f = \rho^{-1}\circ f'\circ\rho$; thus $f\in\Part$.
Actually, $f$ is total, since $\rng(f'\circ\rho) = \rng(\rho\circ
f) \subseteq \rng\rho$. Therefore, $M = \Rec$.
\end{proof}

By the same token:

\begin{theorem}\label{thm:PartIsComplete}
The partial recursive functions \Part\ are complete.
\end{theorem}

\begin{corollary}\label{cor:strong-complete}
The general recursive functions {\rm (\Rec)} and partial recursive
functions {\rm (\Part)} are not strongly equivalent to any of
their submodels or supermodels.
\end{corollary}

\begin{proof}
Non-equivalence to supermodels is just Theorems
\ref{thm:RecIsComplete} and \ref{thm:PartIsComplete}.
Non-equivalence to submodels follows from Lemma
\ref{lem:ProperSubsetBijection}.
\end{proof}

As a corollary of Theorems~\ref{thm:ABC} and
~\ref{thm:RecIsComplete}, we obtain a criterion for
hypercomputation:

\begin{corollary}\label{cor:hyper}
A model $M$, operating over a denumerable domain, is
hypercomputational if there is any bijection under which a proper
subset of $M$ simulates $\Rec$.
\end{corollary}

This justifies the use of the standard comparison method
(Section~\ref{Sec:StandardMethod}) in the particular case of the
recursive functions.

\begin{theorem}\label{thm:TMsimulatesRec}
Turing machines, $\TM$, and the recursive functions, $\Rec$, are
strongly equivalent.
\end{theorem}

\begin{proof}
Since $\Rec$ is complete, it is sufficient, by
Lemma~\ref{lem:CompleteStrongEquiv}, to show that $\Rec \succsim \TM
\succsim_\pi \Rec$, for some bijection $\pi$. Since it is well-known
that $\Rec \succsim \TM$ via G\"{o}delization (e.g.\ \cite[pp.\
208--109]{Jones}), it remains to show that $\TM \succsim_\pi \Rec$,
for some bijection $\pi$. Define (as in \cite[p.\ 131]{Jones}) the
bijection $\pi:\Nat \to \{0,1\}^*$, by
\begin{eqnarray*}
\pi(n) &=& \left\{\begin{array}{ll} \epsilon&~n = 0\\
\begin{array}{l} d~\mbox{s.t. $1d$ is the shortest binary} \\ \mbox{representation of
$n+1$} \end{array} &~\mbox{otherwise}\end{array}\right.
\end{eqnarray*}
For example, $\pi(0,1,2,3,4,5,6,7,\ldots)$ is
$\epsilon,0,1,00,01,10,11,000,\ldots$.

$\TM \succsim_\pi RAM$(Random Access Machine), by \cite[pp.\
131--133]{Jones}; $RAM \supseteq CM$ (Counter Machine), by
\cite[pp.\ 116--118]{Jones}; and $CM \succsim_\iota \Rec$ by
\cite[pp.\ 207--208]{Jones}. We have that $\Rec \succsim \TM
\succsim_{\pi} RAM \succsim_\iota CM \succsim_\iota \Rec$, thus
$\TM$ and $\Rec$ are strongly equivalent.
\end{proof}

Note that the exact definitions of $RAM$ and $CM$ are of no
importance, as they are only intermediates for $\Rec \succsim \TM
\succsim_{\pi} RAM \succsim_\iota CM \succsim_\iota \Rec$.

\begin{theorem}\label{thm:TMComplete}
Turing machines, $\TM$, are complete.
\end{theorem}

\begin{proof}
By Theorem~\ref{thm:TMsimulatesRec}, $\TM$ and $\Rec$ are strongly
equivalent. Since $\Rec$ is complete, it follows, by
Theorem~\ref{thm:StrongEquivComplete}, that $\TM$ is complete.
\end{proof}

\section{Discussion}

There are various directions in which one can extend the work
described above:

\subject{Inductive Domains.} The completeness of the (general and
partial) recursive functions is due to several properties, among
which is the inclusion of a successor function
(Theorem~\ref{thm:RecIsComplete}). The results herein can be
extended to show that computational models operating over other
inductively-defined domains are also complete.

\subject{Intensional Properties of Completeness.} Intuitively, a
properly defined computational model should be complete.  What is,
however, ``properly defined''? One can look for the intensional
properties of a model that guarantee completeness. That is, what
internal definitions that constitute a model (e.g.\ a finite set of
instructions, over a finite alphabet, \ldots) guarantee
completeness.

\subject{Different Domain and Range.} The simulation definition
(Definition~\ref{def:simulation}) naturally extends to models $M:
D^k \to D$ with multiple inputs, by using the same encoding $\rho$
for each input component. See, for example, \cite[p.\
29]{Sommerhalder}.

A more general definition is required for models with distinct input
and output domains. This can be problematic as the following example
illustrates:

\begin{example}\label{exm:RE}
Let \RE\ be the recursively enumerable sets of naturals. We define
infinitely many non-r.e.\ partial predicates $\{h_i\}$, which can be
simulated by \RE. Let
\begin{eqnarray*}
h(n) &=& \left\{
\begin{array}{ll}
0&\mbox{~program $n$ halts uniformly}\\
1&\mbox{~otherwise}
\end{array}\right.\\
h_i(n) &=& \left\{
\begin{array}{ll}
0&~n<i\vee h(n)=0 \\
\bot&\mbox{~otherwise}\;.
\end{array}\right.
\end{eqnarray*}
We have that $\RE \succsim_\rho \RE\cup \{h_i\}$, where
\begin{eqnarray*}
\rho(n) &=& 2n+h(n)\\
h'_i(n) &=& \left\{\begin{array}{ll}
0&~\lfloor n/2\rfloor<i\vee n \bmod 2 = 0\\
\bot&\mbox{~otherwise}
\end{array}\right.\\
\rho(f) &=& \left\{
\begin{array}{ll}
f(\lfloor n/2\rfloor)&~f\in \emph{RE}\\
h'_i(n)&~f=h_i\;.
\end{array}\right.
\end{eqnarray*}
Without loss of generality, we are supposing that $\rho(0)=h(0)=0$.
\end{example}

\subject{Firm comparison.} Comparison by an injective mapping
between domains might be too permissive, as shown in
Example~\ref{exm:RE} above. Accordingly, one may add other
constraints on top of the mapping. For example, adding the
requirement that the ``stronger'' model can distinguish the range of
the mapping. That is, requiring a total function in the ``stronger''
model, whose range is exactly the range of the comparison mapping.

\subject{Multivalued Representations.} It may be useful to allow
several encodings of the same element, as long as there are no two
elements sharing one representation, something injective encodings
disallow. Consider, for example, representing rationals as strings,
where ``1/2'', ``2/4'', ``3/6'', \ldots, could encode the same number.
See, for example, \cite[p.\ 13]{Weihrauch}. To extend the notion of
computational power (Definition~\ref{def:ComputationalPower}) to
handle multivalued representations, we would say that model
$A\succsim B$ if there is a partial surjective function $\eta: \dom
A \to \dom B$ ($\eta(y)=\bot$ iff $y=\bot$), such that there is a
function $f\in A$ for every function $g\in B$, with $\eta(f(x)) =
g(\eta(x))$ for every $x\in \dom \eta$. This follows along the lines
suggested in \cite[p.\ 16]{Weihrauch}. The corresponding definitions
and results need to be extended accordingly.

\subject{Different Cardinalities.}
It may sometimes be unreasonable to insist that the encoding be
injective, since the domain may have elements that are distinct,
but virtually indistinguishable by the programs. For example, a
model may operate over the reals, but treat all numbers $[n:n+1)$
as representations of $n\in \Nat$.

\subject{Effectivity.} A different approach to comparing models over
different domains is to require some manner of effectiveness of the
encoding; see \cite[p.\ 21]{Engeler} and \cite[p.\ 290]{Hennie}, for
example. There are basically two approaches:
\begin{enumerate}
\item One can demand an informal effectiveness:
``The coding is chosen so that it is itself given by an informal
algorithm in the unrestricted sense'' \cite[p.\ 27]{Rogers}.
\item One can require effectiveness of the encoding function via a specific model,
usually Turing machines: ``The Turing-machine characterization is
especially convenient for this purpose. It requires only that the
expressions of the wider classes be expressible as finite strings
in a fixed finite alphabet of basic symbols'' \cite[p.\
28]{Rogers}.
\end{enumerate}
Effectivity is a useful notion; however, it is unsuitable for our
purposes. The first, informal approach is too vague, while the
second can add computational power when dealing with subrecursive
models and is inappropriate when dealing with non-recursive
models.

\subject{Nondeterministic Models.} The computational models we
have investigated are deterministic
(Definition~\ref{def:ComputationalModel}). The corresponding
definitions and results should be extended to nondeterministic
models, as well.

\bibliographystyle{plain}
\bibliography{models1}
\end{document}